\documentclass[twocolumn]{aastex63}
\usepackage{float}
\usepackage{lineno}
\usepackage{graphicx, natbib, color, bm, amsmath, epsfig,hyperref}
\usepackage{lipsum}
\usepackage[normalem]{ulem}

\usepackage{graphicx, natbib, color, bm, amsmath, epsfig}

\newcommand\jcompphys{J.~Comput.~Phys}

\newcommand\revmodphys{Rev.~Mod.~Phys.~}

\newcommand{\K}{\,\rm K}

\newcommand{\mach}{\ensuremath{\mathcal{M_S}}}

\newcommand{\alfmach}{\ensuremath{\mathcal{M_{\rm{A}}}}}
\newcommand{\Ma}{\ensuremath{\mathcal{M_{\rm{A}}}}}
\newcommand{\Ms}{\ensuremath{\mathcal{M_{\rm{s}}}}}

\newcommand{\dbd}[2]{\frac{ \partial #1 }{ \partial #2} }

\newcommand{\p}[1]{#1^\prime }

\definecolor{orange}{rgb}{1.        ,  0.54,  0}

\definecolor{purple}{rgb}{0.5,0.0,0.5} 

\definecolor{meta}{rgb}{0.371,0.617,0.625} 

\newcommand{\dc}[1]{}

\def\Hvec{\ensuremath{{\mathbf H}}}

\def\vvec{\ensuremath{{\bf v}}}

\def\kvec{\ensuremath{{\bf k}}}

\definecolor{pink}{rgb}{1.        ,  0.75294118,  0.79607843}
\definecolor{maroon}{rgb}{0.69019608,  0.18823529,  0.37647059}

\newcommand{\dccnote}[1]{}

\definecolor{gray}{rgb}{0.5,0.5,0.5}

\def\alf{Alfv\' en}

\newcommand{\numvalue}[1]{#1}

\def\ckt{\ensuremath{C_k^{TT}}}
\def\cke{\ensuremath{C_k^{EE}}}
\def\ckb{\ensuremath{C_k^{BB}}}
\def\ckr{\ensuremath{C_k^{\rho\rho}}}
\def\ckv{\ensuremath{C_k^{{vv}}}}
\def\ckh{\ensuremath{C_k^{HH}}}
\def\alphar{\ensuremath{\alpha_{\rho\rho}}}
\def\alphav{\ensuremath{\alpha_{{vv}}}}
\def\alphah{\ensuremath{\alpha_{HH}}}
\def\alphat{\ensuremath{\alpha_{TT}}}
\def\alphae{\ensuremath{\alpha_{EE}}}
\def\alphab{\ensuremath{\alpha_{BB}}}

\def\alphaxx{\ensuremath{\alpha_{XX}}}
\def\alphatot{\ensuremath{\alpha_{\rm{total}}}}
\def\at{\ensuremath{A_{\rm{TT}}}}
\def\ae{\ensuremath{A_{\rm{EE}}}}
\def\ab{\ensuremath{A_{\rm{BB}}}}
\def\axx{\ensuremath{A_{\rm{XX}}}}

\def\rxy{\ensuremath{r^{\rm{XY}}}}
\def\rte{\ensuremath{r^{\rm{TE}}}}
\def\rtb{\ensuremath{r^{\rm{TB}}}}
\def\reb{\ensuremath{r^{\rm{EB}}}}
\def\planckee{-2.42}
\def\planckeeerr{0.02}
\def\planckbberr{0.02}

\def\planckratio{0.53}
\def\planckratioerr{0.01}

\begin{document}

\title{Planck dust polarization power spectra are consistent with strongly
supersonic turbulence}

\begin{abstract}
The polarization of the Cosmic Microwave Background (CMB) is rich in information but obscured by foreground emission from the Milky Way's interstellar medium (ISM).
To uncover relationships between the underlying turbulent ISM and the foreground power spectra,
we simulated a suite of driven, magnetized, turbulent models of the ISM, varying the fluid properties via the sonic Mach number, \mach, and magnetic (Alfv\'en)  Mach number, \alfmach.  
We measure the power spectra of density ($\rho$), velocity ($v$), magnetic field
($H$), 
total projected intensity ($T$), parity-even polarization ($E$),
and parity-odd polarization ($B$).  
We find that the slopes of all six quantities increase with \mach. Most 
increase with \alfmach, while the magnetic field spectrum steepens with
\alfmach.
By comparing spectral slopes of $E$ and $B$ to those measured by Planck, we
infer typical values of \mach\ and \alfmach\ for the ISM.  
 As the fluid velocity increases, $\mach > 4$, the ratio of BB power to EE power increases to approach a constant value near the Planck-observed value of $\sim 0.5$, regardless of the magnetic field strength.
We also examine correlation-coefficients between
projected quantities, and find that \numvalue{$\rte\approx 0.3$}, in agreement with Planck, for appropriate combinations of \mach\ and \alfmach.  Finally, we consider parity-violating correlations \rtb\ and \reb.

\end{abstract}

\author{Kye A. Stalpes}
\affil{Department of Physics, Florida State University, Tallahassee, Florida}
\author{David C. Collins}
\affil{Department of Physics, Florida State University, Tallahassee, Florida}
\author{Kevin Huffenberger}
\affil{Department of Physics, Florida State University, Tallahassee, Florida}
\section{Introduction}

Primordial gravitational waves, generated during inflation and imprinted on the
surface of last scattering, are one of the most exciting sources of
polarization in the Cosmic Microwave Background (CMB).  Their discovery would
give revolutionary evidence for inflation and its mechanism
\citep{Kamionkowski16}.  However,
the \textit{brightest} diffuse sources of polarization in the microwave sky are thermal
dust and synchrotron emission.  These are also exciting signals, because they
reveal the turbulent magnetic field in the interstellar medium (ISM) of our own
Galaxy \citep{Planck15XIX, Kritsuk18,Kim19}.  In order to see the potential inflation signal, we must first
characterize and mitigate the Galactic signal which is at least ten times larger
\citep{Planck18XI}.

Linearly polarized light can be described by the Stokes
$Q,U$ parameters, but
these quantities are coordinate dependent. The coordinate-independent $E,B$
parameters, on the other hand, transform the $Q, U$ signal into parity-even $E$ modes and parity-odd $B$ modes.  The transformation is nonlocal, and size of the kernel depends on the band limit \citep{Rotti19}.  At the surface
of last scattering, cosmological scalar density perturbations produce $E$, while gravitational waves  (cosmological tensor perturbations) are the
only producer of primordial $B$-mode polarization, peaking at degree scales and larger.  As the CMB photons
travels from the surface of last scattering along the line of sight,
gravitational lensing by large scale structure also generates $B$-modes from the scalar $E$ modes.  All of these encode rich information about the physics and cosmology of the Universe.

In the ISM,  $E$ and $B$ (because they are nonlocal) depend on the
geometry of structures and their spatial relationship to the magnetic field
(which sets the polarization directions).  Gravitational, pressure, and magnetic
forces can produce ISM structures with one long and two short dimensions,
so-called filaments, and such filamentary structures are apparent in millimeter
and HI data  \citep{2014ApJ...789...82C, 2016A&A...586A.141P, Clark19} as well
as ISM simulations \citep{deAvillez05,2013A&A...556A.153H}.  A filament will produce predominantly $E$ modes when paired with a magnetic field parallel or perpendicular to the long axis, and a filament will produce predominantly $B$ modes when paired with an oblique magnetic field \citep{Huffenberger20}.  We may thus expect that a sufficiently strong magnetic field (that aligns filaments to its direction) will produce less $B$ signal than $E$.


The Planck satellite measured $E$-mode and $B$-mode power spectra at 353 GHz,
where the signal is dominated by dust emission in the ISM, and found approximate powerlaws
with slopes of  \numvalue{$\alphae =\planckee \pm \planckeeerr$ and
$\alphab=-2.54 \pm \planckbberr$} for a 71\% sky area
\citep{Planck18XI}.  For the same sky area, the ratio of amplitudes of $B$ power
to $E$ power was \numvalue{$\planckratio\pm\planckratioerr$}.  They also found that the scalar temperature $T$-mode and $E$-mode are
correlated with correlation coefficient $r^{TE}=\numvalue{0.36}$, with some scatter but no clear trend with sky region or with angular scale.  Curiously, the $T$-modes and
$B$-modes are correlated with $r^{TB}=\numvalue{0.05}$. This was unexpected as the
correlation of a parity-even $T$-mode and a parity-odd $B$-mode should be zero
on average in systems with no helicity or parity violation, and may indicate
something about the structure of the magnetic field in the solar neighborhood
\citep{Planck18XI}.  \citet{Huffenberger20} pointed out that in a filamentary
picture, a positive $TB$ dust correlation would imply a positive $EB$
correlation.  Such a signature is too small to detect with Planck data alone, but is
amenable to searches assisted with ISM tracers like HI \citep{Cukierman23}.

In addition to searches for inflation, the polarization of the CMB can be used
to hunt for signatures of other beyond-the-standard-model physics.  Cosmic birefringence is the
rotation of the polarization of CMB photons by hypothetical pseudoscalar
fields.  Possibilities include  axion-like particles that may be responsible for
dark matter \citep{Komatsu22}.   \citet{Minami20} employed a strategy to measure the CMB's $EB$ correlation, using the foreground $EB$ correlation to calibrate detector polarization angles.  Based on the CMB $EB$ correlation, they 
report a rotation angle of \numvalue{$\beta=(0.35 \pm 0.14)^\circ$} due to cosmic birefringence, when they assume that the foreground $E$ and $B$ polarization produced by the ISM are uncorrelated.  If instead the sign of the foreground $EB$ is assumed positive (as implied in the filament picture above or for any other reason), their result indeed gets stronger.
\citep[][provide refinements and robustness tests to this approach.]{2020PTEP.2020j3E02M,2022PhRvL.128i1302D,2023JCAP...01..044D}

Simulations of the ISM \citep{Kritsuk18, Kim19} and
theoretical considerations \citep{Caldwell17, Kandel17} have shown that for certain
parameters, magnetohydrodynamic (MHD) turbulence can reproduce the expected $E$-
and $B$-mode power spectra.  That turbulence produces power-law polarization spectra is unsurprising, as $E$ and
$B$ are produced by a combination of quantities that all have power law spectra due to the turbulent cascade:
density \citep{Beresnyak05, Collins12}, velocity
\citep{Kolmogorov41,Goldreich95}  and magnetic
field \citep{Grete23}.

In this work, we characterize how the polarization spectra depend on the MHD fluid parameters.  We perform idealized simulations of MHD turbulence in order to 
characterize the $E$ and $B$ spectra and their correlations.  We examine the
power spectra of 3d fluid quantities, density $\rho$, velocity $\vvec$, and magnetic
field $\Hvec$; as well as 2d projected observable quantities, $T$, $E$, and $B$.  
From these results, we infer averaged properties of the ISM from the Planck measurements. 

We organize our paper as follows.  We describe the methods for simulations and
analysis in Section \ref{sec.methods}.  We present results, beginning with a short overview, 
in Section \ref{sec.overview}.  
There we show power spectra and  slopes for both fluid (Section
\ref{sec.primitives}) and projected quantities (Section \ref{sec.ebslopes}). We
present ratios of $T$, $E$, and $B$ and thier correlation coefficients (Section \ref{sec.ratio}).  Using the
measured quantities from Planck, we posit typical values for the  ISM's sonic and \alf\ Mach
numbers (Section \ref{sec.linear}).  We briefly contrast these findings with
projections parallel to the mean magnetic field (Section \ref{sec.xhat}).  
We discuss the relation between these findings and our filamentary model
(Section \ref{sec.discussion}).
We conclude in Section \ref{sec.conclusions}.

\begin{figure} 
\includegraphics[width=0.95\linewidth]{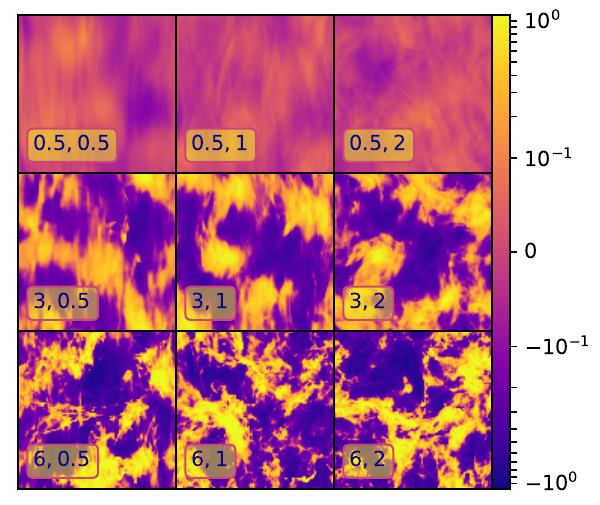}
\includegraphics[width=0.95\linewidth]{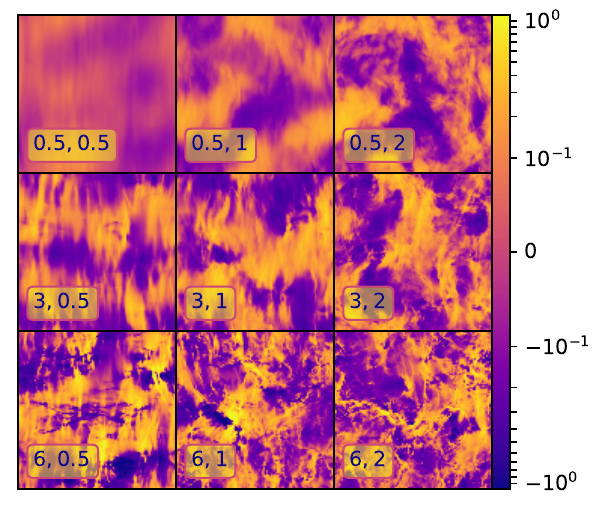}
\includegraphics[width=0.95\linewidth]{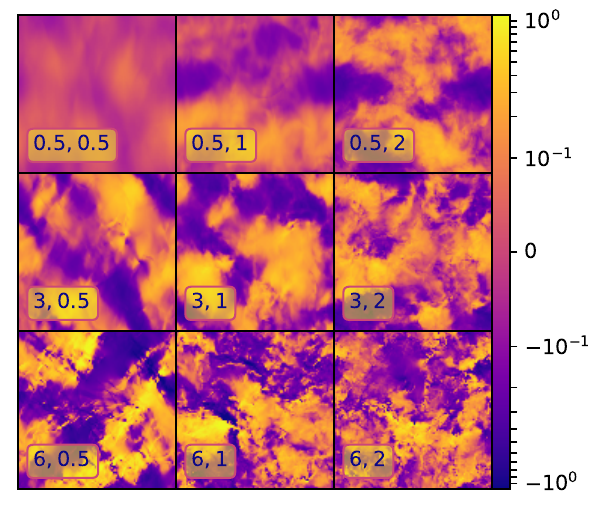}
\caption[ ]{Images of $\ln(T)$ (top), $E$ (middle), and $B$ (bottom).  The mean magnetic field points up. 
The colorbar is a symmetric logarithm.  
 In each $3 \times 3$ panel, simulations with weaker magnetic fields (higher $\Ma$) are to the right and simulations with faster fluid flow (higher $\Ms$) are to the bottom.  Nine of the 21 total simulations are shown, with the targeted $(\Ms, \Ma)$ indicated in the box.
}
\label{fig.proj}  \end{figure}

\begin{figure}
\includegraphics[width=\linewidth]{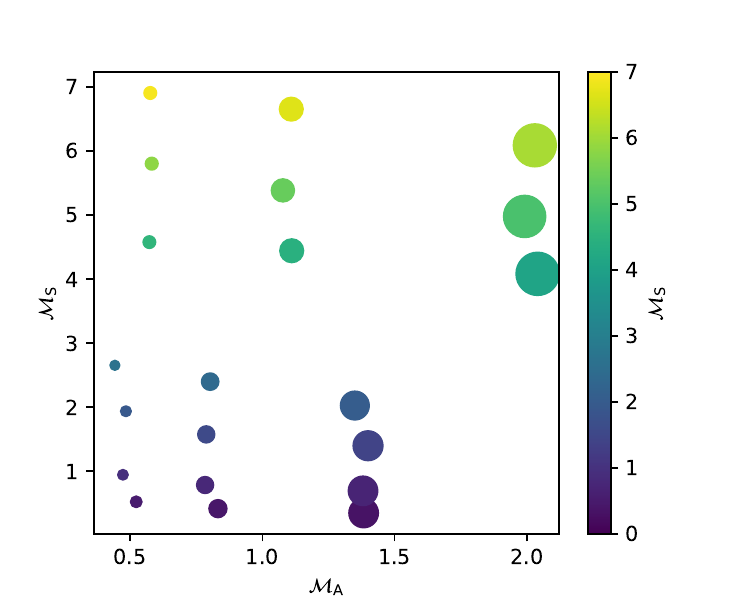}
\caption{Legend of sonic and Alf\'en Mach numbers, \mach\ and \alfmach, for each simulation.  As with all future plots,
color denotes \mach, with
blue-to-yellow indicating indicating increasing  sonic
Mach numbers (and faster rms fluid velocity).
Marker size
denotes \alfmach\ (increasing size indicates increasing \Ma, and weaker influence of the magnetic field), 
}
\label{fig.nominal}
\end{figure}
\section{Methods}
\label{sec.methods}

We perform a suite of idealized simulations of the intersteller medium.  From the 3d simulation boxes, we compute temperature and polarization images, and then compute spectra of 3d and 2d physical quantities.
Ideal MHD has three independant quantities: density, $\rho$, velocity,
$\vvec$, and magnetic field, $\Hvec$.  With the ansatz that turbulence
dictates the primary behavior in $E$- and $B$-modes, we examine the variation
in power spectra in all six quantities of interest: the three fluid
quantities, $\rho$, $\vvec$, and $\Hvec$; and the three projected quantities, $T$, $E$,
and $B$.

\subsection{Simulations}

The simulations utilize the open source code Enzo \citep{Wang09,Bryan14} to
solve the Eulerian equations of ideal magnetohydrodynamics (MHD).   We used the \citet{Dedner02} divergence cleaning scheme with a piecewise linear
reconstruction and HLLD Riemann solver \citep{Mignone07}.  With an adiabatic index $\gamma -1 = 10^{-3}$, we achieve a reasonable approximation to an isothermal equation of state.

Ideal isothermal MHD is scale free.
The dynamics depend on only two parameters, and we choose
the sonic Mach number \mach, and the \alf\ Mach number, \alfmach:
\begin{align}
\mach &= v_{\rm rms}/c_s \label{eq:Ms}\\
\alfmach &= v_{\rm rms}/v_A \label{eq:Ma}\\
v_A &= H/\sqrt{4\pi \rho} \label{eq:Va}
\end{align}
where $\rho$ is the mean density, $v_A$ is the \alf\ velocity, $c_s$ is the
speed of sound, and $H$ is the mean magnetic field strength.  Stronger field means bigger $v_A$ and thus smaller $\alfmach$.  In these simulations we drive turbulence in periodic boxes, altering the sonic and \alf\ Mach numbers, and explore the resulting power spectra and correlations.

\def\tdyn{\ensuremath{t_{\rm{dyn}}}}
The simulation boxes are periodic, use a $512^3$-zone resolution, and start with uniform density.  To generate turbulence, we drive the fluid with a stochastic forcing that adds a random acceleration pattern 
 to the velocity field in a way that keeps the energy injection rate
constant \citep{MacLow04, Federrath10}.  The driving is applied for ten
dynamical times, $\tdyn=L_0/\mach$, where $L_0$ is the pattern size, half the box size, and \mach\ is the
r.m.s.\ velocity.  
The first 5 \tdyn\ are ignored, and used only to establish the fully developed
turbulence. The remaining 5 \tdyn\ are used for analysis.  The force is 
distributed as a Gaussian in each component, with power only on large spatial scales, $k/k_{\rm min}\in[1,2]$.  The forcing pattern is evolved in
time with an  Ornstein-Uhlenbeck (OU) process \citep{Federrath10}.  This means
that the driving pattern retains only 1/e of its correlation after \tdyn.
The input power is
split between compressible and solenoidal modes such that the ratio of
compressive to solenoidal amplitude is 2/3 ($\zeta$ = 1/2 in Equation 6 of
\citet{Federrath10}).  It is anticipated in the inertial range for compressible
turbulence, the natural ratio is 2/3 \citep{Kritsuk07}. {We have checked that the driving field's helicity ($\int dV\  \mathbf{v} \cdot (\nabla \times \mathbf{v})$) is zero to machine precision, so we do not expect the driving to introduce parity violation.}

In practice, the mean density and sound speed are both set to unity, and the
Mach number is controlled by setting the energy injection rate, $\dot{E}\propto
\mach^3/L$.

Throughout the work we define \mach\ and \alfmach\ with the 3D velocity dispersion,
$v_{\rm rms}^2= \langle v_x^2\rangle + \langle v_y^2\rangle + \langle
v_z^2\rangle$. In contrast to the 3D Mach numbers, the 1D Mach number (which enters the Maxwellian velocity
distribution \citep{Rabatin23b} and is more accessible from the ground) is found
by assuming isotropy and dividing $v_{\rm rms}$ by $\sqrt{3}$. 

To probe the parameter space, we target nominal sonic Mach numbers $$\mach = 0.5, 1, 2, 3, 4, 5, 6$$ and
nominal Alfven Mach numbers $$\alfmach = 0.5, 1, 2$$ across the suite of 21
simulations.  The actual \mach\ and \alfmach\ realized by the simulations differ slightly from these nominal
values.

Figure \ref{fig.proj} shows $\ln(T)$, $E$, and $B$ (defined precisely in the
following section). Each figure shows three target sonic Mach numbers, (0.5, 3, 6) and
\alf\ Mach numbers (0.5, 1, 2). The colorbar is a symmetric logarithm.  
Structure changes can clearly be seen as \mach\ and \alfmach\ increase.  
This is due to the fact
that the dissipation scale decreases as the mean kinetic energy increases.

\label{sec.colors}

Figure 
\ref{fig.nominal} shows a legend of the achieved \mach\ and \alfmach\ for each
simulation.
Because it is challenging to predict the forcing to achieve particular Mach numbers, the measured values of \mach\ and
\alfmach\ differ slightly from the nominal values listed above.  
This figure also serves as a legend for the remaining figures in the work, with
color denoting sonic Mach number, with dark blue-to-yellow
corresponding to achieved \mach, ranging from subsonic ($\mach<1$) to supersonic ($\mach \sim 7$). 

Marker size
increases with \alfmach, so larger markers denote weaker magnetic field
strength.
In the power-spectra plots, we will use linestyles to denote \alfmach, with dotted lines for $0.4 < \alfmach < 0.7$, dashed lines for $0.7<\alfmach<1.2$, and solid lines for $1.2<\alfmach<2.2$.

\subsection{Projection to dust temperature and polarization}

The polarization we are focusing on here comes from elongated dust grains that
rotate around the local magnetic field, with their long axis perpendicular to
the field direction.  
We make several simplifying assumptions: the dust-to-gas ratio is constant and uniform, the
dust grains perfectly align with the magnetic field, the cloud is
optically thin, the dust temperature is the same as the  gas temperature (which
are both constant and uniform), and there is only one dust species. The boxes are scale-free and do not correspond to any particular physical size.

We focus most of our attention on projections perpendicular to the
mean magnetic field.  This is because observations oblique to the mean magnetic field
are more likely than along the mean magnetic field, as
 the solid angle for to vectors to be nearly aligned is much smaller than it is for
them to be nearly perpendicular.  
Of course, line-of-sight alignment may exist over some portion of the sky, and
the true picture is a mixture of angles. We will start with the more observable
case, and return to discuss
parallel projections in section \ref{sec.xhat}.

From the assumption of optically thin dust, the $T$-mode is simply proportional
to the column density,
\begin{align}
T = \int \rho dz.
\end{align}
To compute $E$ and $B$, we first compute Stokes parameters $Q$ and $U$, which
are closely related to the observable quantities.  These are
\begin{align}
	Q &= \int \rho \cos{2 \psi} \cos^2{\gamma} dz\label{eqn.Q}\\
	U &= \int \rho \sin{2 \psi} \cos^2{\gamma} dz \label{eqn.U},
\end{align}
where $\psi$ is the angle the field makes in the plane of the sky relative to
horizontal, and $\gamma$
is the angle between the magnetic field and the plane of the sky \citep{Bohren98,
Fiege00b}.  For projections
along the $\hat{z}$-axis line-of-sight, and choosing $\hat{x}$ as the horizontal direction, this
gives
\begin{align}
Q &= \int \rho \frac{H_x^2-H_y^2}{H_x^2+H_y^2+H_z^2} dz\\
U &= \int \rho \frac{2 H_y H_x}{H_x^2+H_y^2+H_z^2} dz.
\end{align} 
In the flat-sky approximation, the coordinate-invariant quantities $E$ and $B$ are then found as
\begin{align}
\tilde{E} + i \tilde{B} &= \left( \tilde{Q} + i \tilde{U}\right) e^{-2i
\theta_k}, \label{eqn.EB}
\end{align}
where $\tilde{E}$ denotes the Fourier transform of $E$, and $\cos \theta_k =
{k_x}/{({k_x^2+k_y^2})^{1/2}}$ is the angle in Fourier space \citep{Kamionkowski16}.

\subsection{Power spectra}
\label{sec.spectra}

We compute the \emph{average} power spectra of all quantities by averaging over a shell or annulus in Fourier space:
\begin{align}
  C_k^{XY} = \frac{1}{\Delta V_k} \int_{||\kvec'|-k| < \Delta k} d^D \kvec' \tilde{X}(\kvec') \tilde{Y}(\kvec')
  \label{eqn.powerspectrum}
\end{align}
where $\tilde{X}$ and $\tilde{Y}$ are Fourier transforms of fluid quantities
($\rho$, $\mathbf{v}$, and $\mathbf{H}$, whence dimension $D=3$)
or projected quantities ($T$, $E$, and $B$, whence $D=2$).  $\Delta V_k$ is the volume
of a shell at $k$, which has thickness matched to the resolution of the Fourier
grid, $\Delta k = k_{\rm min} = 2\pi/L$. 
Because the box is scale-free, the wavenumbers $k$ do not correspond to any particular angular scale or multipole on the sky.

For vector quantities the product $XY$ is replaced with the vector dot product,
e.g.
\begin{align}
	C_k^{vv} = C_k^{v_x v_x}+C_k^{v_y v_y}+ C_k^{v_z v_z},
\end{align}
and a similar expression for the magnetic field.

Quite often the turbulence literature employs the contribution to the \emph{total} power in a shell, which omits the shell
volume, $\Delta V_k$, in
Equation \ref{eqn.powerspectrum}, while the cosmology literature uses the average power in the shell for the CMB and large-scale structure.  The convention in the turbulence literature is due
to the relationship between the power spectrum and the total energy in the
system \citep[e.g.][]{Pope00}.  The slope of \emph{total}-power spectrum can be recovered
from the \emph{average}-power spectra presented here as $\alpha_{\rm{total}}=\alpha_{XX}+2$ in three-dimensions.  This is most
apparent when examining the velocity spectrum: the \emph{total} power in the
traditional Kolmogorov cascade is
$\alpha_{\rm{total}}=-5/3$, while the \emph{average} value is $-11/3$.  We use
the \emph{average} spectrum throughout to connect with the Planck-measured CMB power spectra.

Each spectra can be broken into three regimes.  At large scales \numvalue{($k<
k_{\rm{drive}}$)}, the driving of the turbulence dominates these spectra, which 
depends on the details of the simulator's particular setup.  At small scales \numvalue{($k>
k_{\rm{diss}}$)}, the spectra is dominated by numerical dissipation.  In between, in the so-called inertial range where we are most interested in the behavior, the spectra are set by the nonlinear dynamics of the system.  
Empirically, we use $k_{\rm{drive}}=4 k_{\rm{min}}$ and $k_{\rm{diss}}=25
k_{\rm{min}}$, as this range captures the nearly-powerlaw section of each of the spectra (visible in Fig.\ \ref{fig.primspectra} and Fig.\ \ref{fig.TEB_spectra}).
In this range, we fit the
spectra to the form
\begin{align}
	C_k^{XX} = \axx k^{\alphaxx}.
	\label{eqn.fit}
\end{align}
To estimate uncertainties on the spectral slopes, we computed $\alphaxx$ for every simulation timestep, also varying $k_{\rm{drive}}\in [3,4,5] k_{\rm{min}}$ and $k_{\rm{diss}} \in
[25,26,27,28]k_{\rm{min}}$, and took the standard deviation of the collection.

\begin{figure}
	\includegraphics[width=\linewidth]{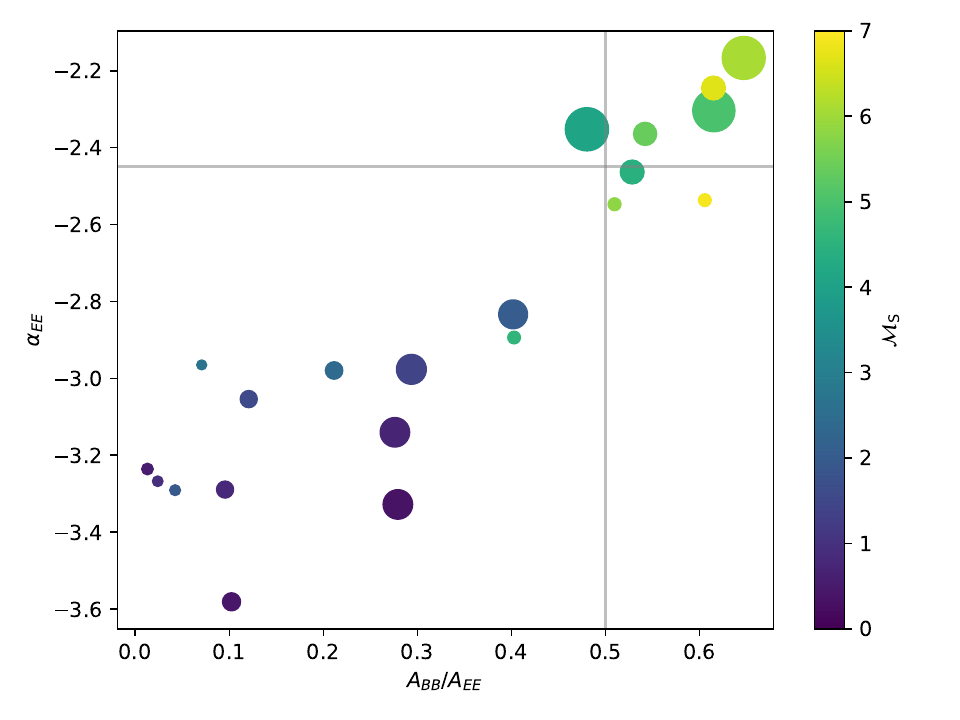}
\caption{A summary of the slope of $E$-mode vs. the ratio of amplitudes,
$\ab/\ae$.  The grey lines denote the approximate values from \numvalue{\citet{Planck18XI}:
$\alphae=-2.42$, $\ab/\ae=0.5$}.  Color and marker size are as described in
Section \ref{sec.colors}}
\label{fig.summary}
\end{figure}
\section{Results}
\label{sec.overview}

Figure \ref{fig.summary} gives an overview of the main result for foreground
polarization.  The vertical axis shows $\alphae$ while the horizontal shows
$\ab/\ae$ (the ratio of the fit amplitudes).  Grey lines indicate the
Planck-measured values.  The color shows the sonic Mach number \mach\ and the size
shows \alfmach\ (as in Fig.~\ref{fig.nominal}).  As discussed below, as \mach\
increases above 4, the ratio  $\ab/\ae$ increases to sit in the range [0.4,0.7],
compared to the Planck value near 0.5, and $\alphae$ becomes shallower to a
range $[-2.6,-2.2]$, compared to the Planck value of \planckee.

\begin{figure*} \begin{center}
	\includegraphics[width=\textwidth]{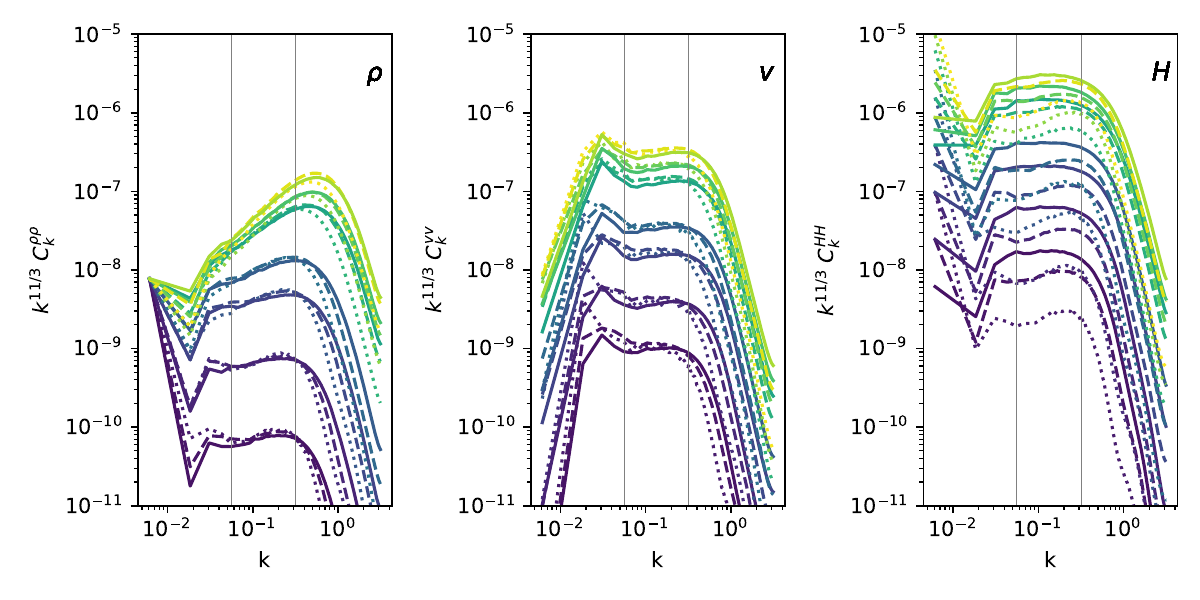}
\caption[ ]{ 
Power spectra of density ($C_k^{\rho\rho}$), velocity
($C_k^{vv}$) and magnetic field strength ($C_k^{HH}$) (left to right,
respectively.) 
All plots have been compensated to the Kolmogorov value of 11/3.
Color (blue-yellow)
denotes increasing sonic Mach number, while line style denotes \alfmach\ as described
in Section \ref{sec.colors}: dotted lines for $0.4 < \alfmach < 0.7$, dashed lines for $0.7<\alfmach<1.2$, and solid lines for $1.2<\alfmach<2.2$.  Density slope depends heavily on sonic Mach number,
while velocity and magnetic slopes do not.
}
\label{fig.primspectra} \end{center} \end{figure*}
\begin{figure*}\ \begin{center}
	\includegraphics[width=\textwidth]{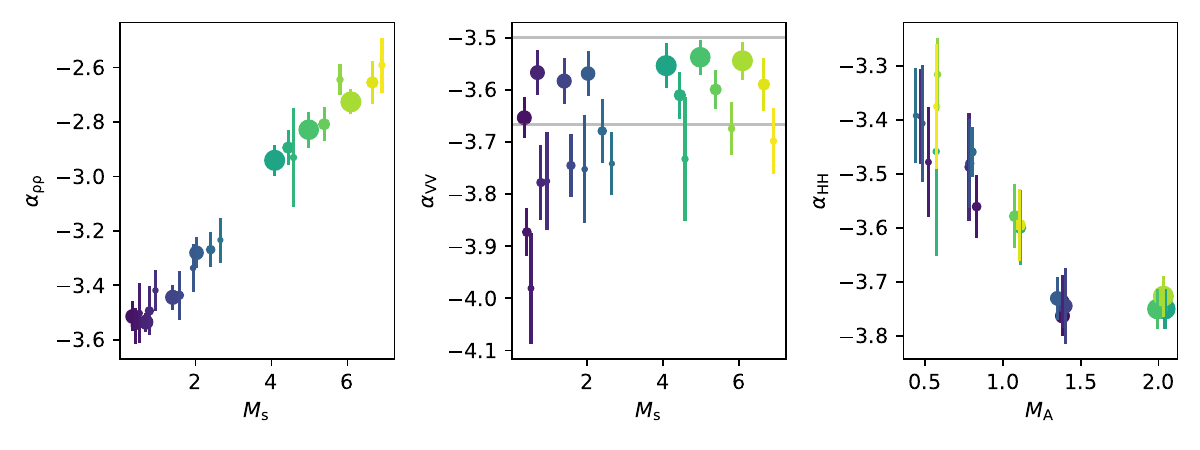}
\caption[ ]{ Slopes 
 of fluid quantities ($\rho$, $\vvec$, and $\Hvec$).
 The first two are relative to \mach\ 
 and \alfmach\ is relative to \alphah. Color denotes \mach\ while size denotes \alfmach\ as described in
 Section \ref{sec.colors}.
}
\label{fig.primslopes} \end{center} \end{figure*}

\begin{figure*} \begin{center}
\includegraphics[width=\textwidth]{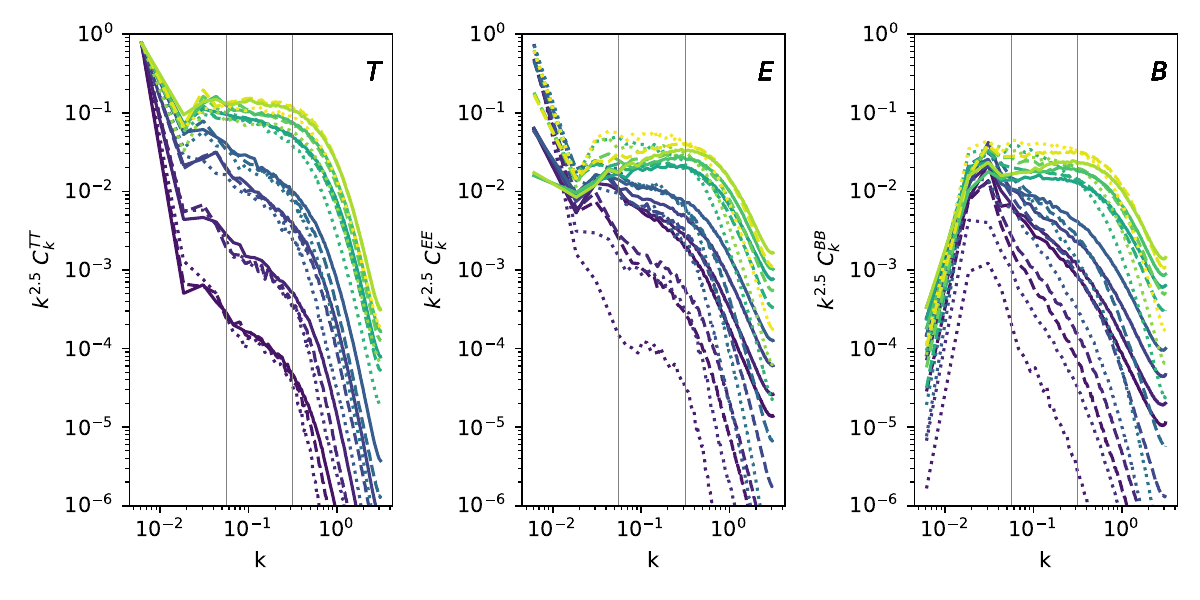}
\caption[ ]{Spectra of projected quantities, $C_k^{TT}$, $C_k^{EE}$, and
$C_k^{BB}$.  Each spectra has been compensated to $k^{2.5}$ to emphasize the
variation with simulation.  Color denotes \mach\ and linestyle denoste
\alfmach, as described in Section \ref{sec.colors}: dotted lines for $0.4 < \alfmach < 0.7$, dashed lines for $0.7<\alfmach<1.2$, and solid lines for $1.2<\alfmach<2.2$.
}
\label{fig.TEB_spectra} \end{center} \end{figure*}
\begin{figure*} \begin{center}

	\includegraphics[width=\textwidth]{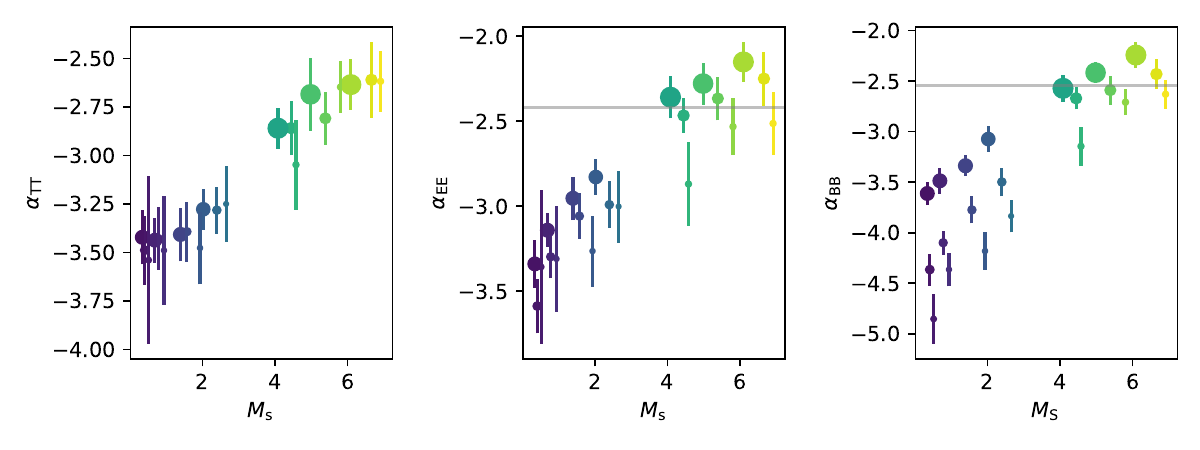}
\caption[ ]{ Slopes of intensity and polarization power spectra versus sonic Mach number.  The horizontal axis and color denote \mach\ while size denotes \alfmach\ as described in
 Section \ref{sec.colors}. Grey lines denote the values observed by Planck}
	\label{fig.TEB_slopes} \end{center} \end{figure*}
\subsection{Fluid Power Spectra}
\label{sec.primitives}

Figures \ref{fig.primspectra}  displays the spectra for the fluid quantities,
\ckr\ (density, left), \ckv  (velocity, center) and
\ckh\
(magnetic field, right). Figure  \ref{fig.primslopes} shows the slopes of those
spectra, $\alphar$ (left), $\alphav$  (center) and $\alphah$ (right).
The spectra are compensated by $k^{11/3}$ to emphasize variations relative to the average Kolmogorov slope value, which would be flat in this plot.
The plot style is described in Section \ref{sec.colors}; color (blue
to yellow) denotes increasing \mach, while line style denotes \alfmach.
Figure \ref{fig.primslopes} shows the slopes, \alphaxx, with color denoting
\mach\ and point size increasing with \alfmach.
We will discuss each in turn.

\paragraph{Density} Beginning with density (first panels in Figures \ref{fig.primspectra}  and
\ref{fig.primslopes}) we find that both slope and amplitude are
increasing functions of \mach. 
The increase in amplitude is to be expected, as the variance in 
density is linearly proportional to the variance in velocity
\citep{Padoan97}, i.e. 
\begin{align}
\sigma^2_\rho = b^2 \mach^2,
\end{align}
where $b\in[1/3,1]$ \citep{Federrath08}.  By the Plancherel theorem the variance
is equal to the area under the power
spectrum,
so it is expected that the amplitude of the density power spectra will
increase with sonic Mach number.  
The slope becomes shallower in an almost linear way from $\alphar\sim-3.5$ to $\alphar \sim -2.5$ as
\mach\ increases from 0.5 to 7.  
As \mach\ increases, the typical shock velocity also increases, which gives rise to enhanced structure formation by way of fluid instabilities such as the Richtmyer-Meshkov instability \citep{Richtmyer60,Meshkov72}.  This enhanced power flattens the spectrum.
This
behavior 
has been seen before \citep{Beresnyak05, Collins12}.

Neither slope, $\alpha_{\rho\rho}$ nor amplitude, $A_\rho$ vary with \alf\ Mach
number.  This is not particularly surprising, as the continuity equation which
determines density only contains density and velocity:
\begin{align}
\dbd{\rho}{t} + \nabla \cdot \vvec \rho = 0,
\end{align}
so the density is determined by velocity.  
Specifically, we find that $\alphar$ is determined by the r.m.s.\ velocity alone. However, as we will discuss in the next section, the velocity spectrum
does not evolve with sonic Mach number in the same manner as the density
spectrum, indicating that the shape of the density spectrum is not a direct result of the shape of the velocity spectrum.

\paragraph{Velocity} The velocity field (second panels of Figures \ref{fig.primspectra} and 
\ref{fig.primslopes}) can be seen to vary jointly with \alfmach\ and
\mach.  In incompressible hydrodynamical turbulence, the
expectation is that the slope has the (average) Kolmogorov value of $-11/3 \approx -3.7$.  These
simulations are not imcompressible, but highly compressible and magnetized.  For supersonic
hydrodynamical turbulence, one expects a value of $\alphav=-4$ or
($\alphatot=-2$,
\citealt{Kritsuk07}.)
 For incompressible magnetized turbulence,
 the spectral scaling has been debated, with some authors expecting a value of $\alphav=-5/2$ \citep[\alphatot=$-3/2$][]{Iroshnikov64, Boldyrev06})
and some expecting a value of $-11/3$ 
\citep[\alphatot=$-5/3$][]{Beresnyak11}.  For a recent overview, see \citet{Schekochihin22}.  
There is not a theory that combines compressibility and magnetization that is
appropriate for the simulations presented here,  and in
our case we see some resemblance to all of the above.
The simulation 
that most closely  approaches the un-magnetized incompressible assumption of the
Kolmogorov cascade has \mach$=0.5$ and \alfmach$=2$, which does have a slope of $-3.6$. For low \mach, the slope $\alphav$
steepens from $-3.6$ to $-3.9$ as the field increases (dot size shrinks).  
Once supersonic, the slope of the velocity 
does not vary much with \mach, but does steepen with increasing field strength.
For low field strength, the slope is around around $\alphav=-3.5$
($\alphatot=-1.5$).
Also shown in the figure are horizontal lines showing the fiducial values of $\alphav=-5/2$ ($\alphatot=-3/2$) and $\alphav=-11/3$ ($\alphatot=-5/3$).

\paragraph{Magnetic field} The final fluid quantity is magnetic field, $\Hvec$.  Spectra are plotted in the last panel of Figure
\ref{fig.primspectra}, and slopes are plotted in the last panel of Figure \ref{fig.primslopes}).    Here the magnetic slope,
$\alphah$ is plotted against \alf\ Mach number rather than sonic Mach number.
It can be seen that the slope of the magnetic field, $\alphah$, does not
depend strongly on \mach, as points with similar color cluster around the same
value, but does decrease nearly linearly for decreasing magnetic field strength.
For the weakly magnetized runs, the slope is $\alphah=-3.75$, and it becomes shallower to $-3.3$ for
the strongly magnetized runs.   

This behavior is likely the compressible analog of the transition from  weak
turbulence, where magnetic fluctuations are smaller than the mean, to strong
turbulence, where the fluctuations are large compared to
the mean magnetic field \citep[see][for an excellent review]{Schekochihin22}.  In incompressible simulations, the spectrum is observed
to steepen continually from $\alphatot=-3/2$ to $\alphatot=-2$ as the turbulence
move from weak to strong \citep{Perez08}.  This is similar to the monotonic
steepening in $\alphah$ as we increase \alfmach, which seems to level off above
$\alfmach\sim1.5$.  In addition to the transition from weak to strong turbulence as
\alfmach\ increases, in our simulations we are also transitioning to shock
dominated turbulence, which complicates the picture relative to the incompressible work.

\subsection{Projected Intensity and Polarization Power Spectra}
\label{sec.ebslopes}

Figures \ref{fig.TEB_spectra} and \ref{fig.TEB_slopes} show the spectra and
slopes
for projected quantities, $T$, $E$ and $B$.  Color denotes \alfmach\
(blue-to-yellow denotes increasing \mach) and line style denotes \alfmach\ 
(solid, dashed, and dotted denoting increasing value) as described in
Section \ref{sec.colors}.  All projected spectra have been compensated so that
slope of $\alpha = -2.5$ would appear flat and horizontal.  

\paragraph{Intensity/Temperature} The $T$ spectrum, $\ckt$, can be seen in the first panel of Figures \ref{fig.TEB_spectra}
and its slope, $\alphat$, in the first panel of Figure
\ref{fig.TEB_slopes}.
The $T$ spectrum has a slope that is nearly identical to the $\rho$ spectrum
(though they appear different due to the difference in compensation).
This is expected as in this model, $T$ is simply the projection of
$\rho$.  By the slice-projection theorem, a quantity $q(x,y,z)$, its
projection $Q(x,y)$, and their Fourier transforms $\hat{q}(k_x,k_y,k_z)$ and
$\hat{Q}(k_x,k_y)$ are related as 
\begin{align}
\hat{Q}(k_x,k_y) = \hat{q}(k_x,k_y,k_z=0).
\end{align}
That is, the transform of the projection is the zero mode of the transform along
the projection axis.  Thus one can reasonably expect $T$ and $\rho$ to have the
same \emph{average} power spectra, provided the field is isotropic.
Thus, $T$ power spectral slopes and amplitudes
should also depend primarily on \mach\ in the same linear fashion as
$\alpha_\rho$.  The match is not exact due to the fundamentally anisotropic
nature of our simulations' mean magnetic fields, but quite similar.

\paragraph{E-mode}The even-parity $E$ spectrum, $\cke$, can be seen in the
second panel of Figures \ref{fig.TEB_spectra} and its slope, $\alphae$, 
in the second panel of Figure \ref{fig.TEB_slopes}.  The grey line in Figure
\ref{fig.TEB_slopes} shows the observed value of $\planckee$.
It can be seen that \alphae\ depends on \mach\ nearly linearly,
and \alfmach\ somewhat.   As the sonic Mach number increases, we find that $\alpha_{EE}$ get shallower from $-3.5$ for $\mach=0.5$ to $-2.3$ for $\mach=6$.  Increasing magnetic field (decreasing \alfmach)
steepens the $\alpha_{EE}$ slope.  Simulations with $\mach\ge 4$ are needed to reproduce the measured Planck $\alpha_{EE}$.  Simulations with gas velocities that are too slow yield slopes that are too steep.  The amplitude of the $E$ power increases for decreasing field when
\mach\ is low, but is relatively immune to both \mach\ and \alfmach\ for
supersonic runs.

Interpreting \alphae\ (and \alphab, next section) is tricky, since unlike the previous quantities, we lack even an adjacent theory about its behavior in a turbulent medium.  It set by a combination of the geometry of density and magnetic structures, convolved with a kernel \citep{Rotti19}.  We revisit this interpretation in Section \ref{sec.discussion}.

\paragraph{B-mode} The odd-parity $B$ spectrum, \ckb, can be seen in the right
panel of Figure \ref{fig.TEB_spectra}, and its slope, \alphab, can be seen in
the right panel of Figure \ref{fig.TEB_slopes}.
The slope depends on \mach\ like $\alpha_{EE}$ but has a stronger dependence on \alfmach, particularly when \alfmach\ is small (strong field). For the weakly
magnetized runs, $\alphab$ ranges from $-3.5$ to $-2.1$.   

Compared to the the other projected slopes, there is a strong steepening of the slope \alphab\ at all \mach\ as \alfmach\ decreases.  That is, stronger magnetic fields 
result in steeper power spectra in the $B$ mode.  We may interpret this
reduction in power as a stiffening of the filamentary structure, which cause
the field and filament to more likely align on small scales.  
{We revisit this interpretation in the discussion.}

\begin{figure}
\includegraphics[width=\linewidth]{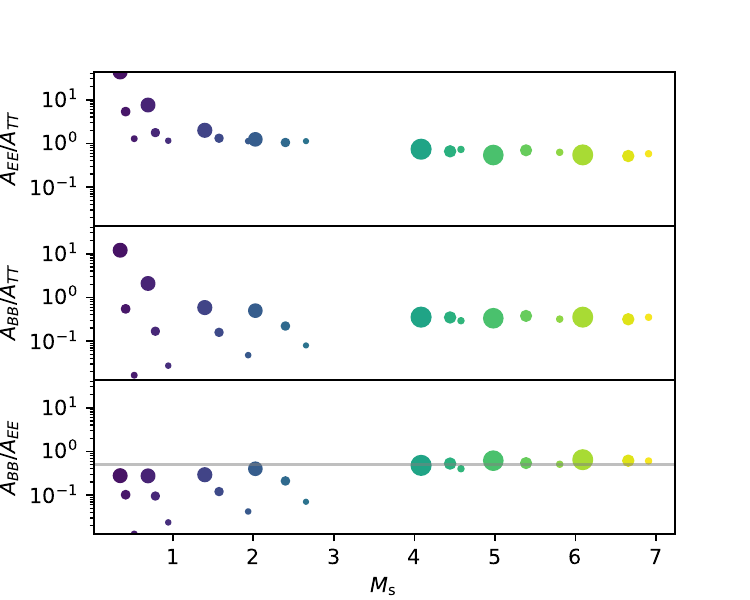}
\caption{Ratios of amplitudes converge on \numvalue{(0.62, 0.34, 0.55)} for
$E/T$, $B/T$, and $B/E$, respectively. Colors are described in Section
\ref{sec.colors}.  A value of 0.5 is seen as the grey line in each row.}
\label{fig.ratio}
\end{figure}
\begin{figure*}
\includegraphics[width=\textwidth]{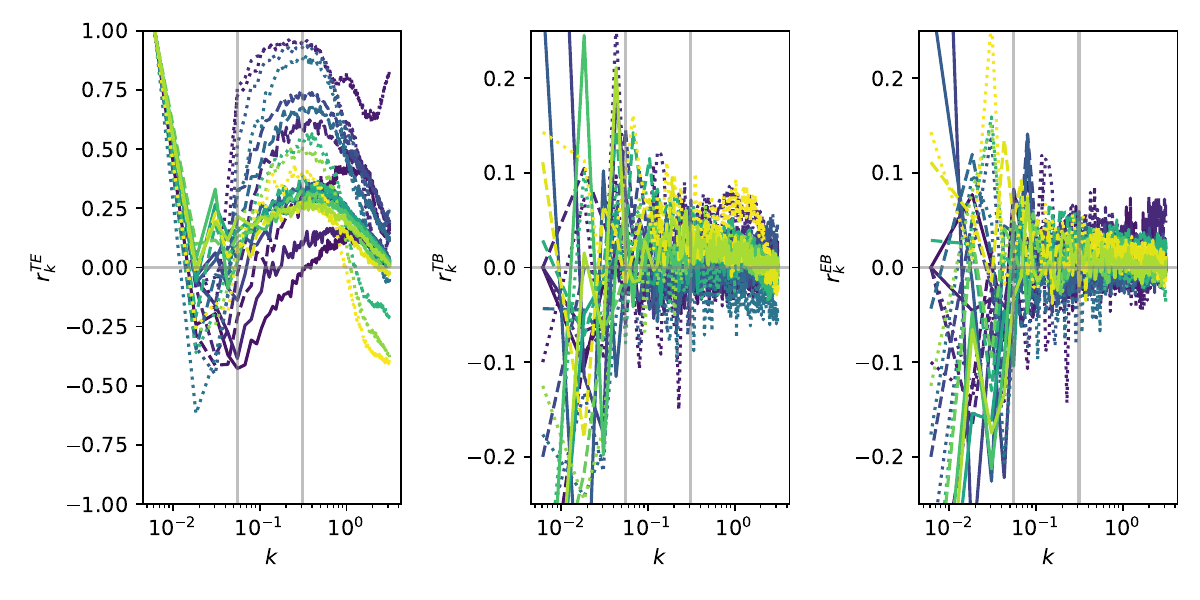}
\includegraphics[width=\textwidth]{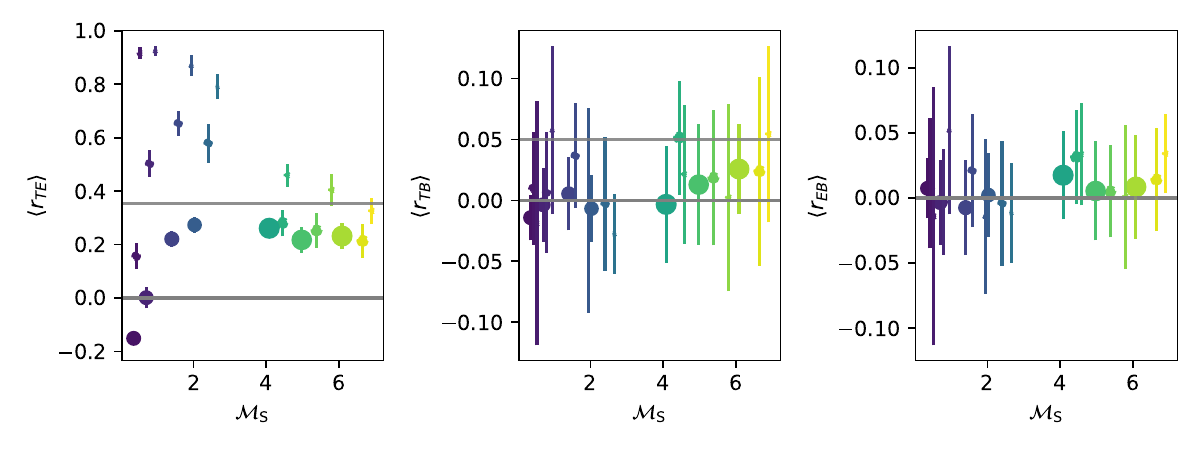}
\caption{(\emph{Top}) Correlation ratio $r_{xy}$ for TE, TB, and EB spectra.
Vertical bars show the analysis window.   (\emph{Bottom}) Average $r_{XY}$ in the analysis window versus
sonic Mach number, $\mach$, compared to Planck-measured values.  In the top plots, we use dotted lines for $0.4 < \alfmach < 0.7$ (smallest point size at bottom), dashed lines for $0.7<\alfmach<1.2$, and solid lines for $1.2<\alfmach<2.2$ (largest point size at bottom).  See text for cautions about interpreting the $\langle r_{TB} \rangle$ and $\langle r_{TE} \rangle$.
}
\label{fig.pearson}
\end{figure*}

\subsection{Power Amplitude Ratios}
\label{sec.ratio}

The ratio of $B$ power to $E$ power is interesting because the observed value of
$\ab/\ae=\planckratio$ was unexpected.  We can also examine the
ratio of of $E$ and $B$ relative to total power, $\ae/\at$ and $\ab/\at$, though
our model lacks a detailed treatment of the dust polarization fraction. This
implies that these ratios cannot be directly compared to observed values, though
the trend with \mach\ and \alfmach\ can be measured.  The unknown polarization fraction
cancels out in the $\ab/\ae$ ratio, so this can be directly compared to the Planck value.

Figure \ref{fig.ratio} shows the ratio of fit amplitudes, $\ae/\at$, $\ab/\at$,
and $\ab/\ae$ versus the sonic Mach number in each of the three panels.  
The third panel also shows the observed value as a grey line.
The runs with slower velocities show the most $E$ modes, as the low velocity causes
the field to have larger impact on the morphology, and more filamentary
structures that align magnetic field and density are observed.  This can be seen
in the projections in Figure \ref{fig.proj}.  
For higher velocity ($\mach \ge 4$), the amplitude ratios depend little on either the fluid velocity or the magnetic field.  
For $\mach>4$, the ratio tends toward $A^{BB}/A^{EE}=0.55\pm0.07$,  near to the
observed ratio of Planck.
Thus it may be that that the ratio of $B$ to $E$ observed by Planck is a natural consequence of
compressive turbulence, and that $\ab/\ae=\planckratio$ simply because the flow
is hypersonic and magnetized, which naturally gives this value.

The study by \citet{Caldwell17} used a linearization of the MHD equations to predict the amplitude ratio but had trouble reproducing the observed value.  They postulate that the reason is the lack of nonlinearity in their treatment.  Here we tend to agree, as our simulations become more nonlinear the observed slope is recovered.

\subsection{Cross-correlations}
Figure \ref{fig.pearson} shows the correlation coefficient spectra,
\begin{equation}
\label{eq.r}
\rxy_k = {C_k^{XY}}/{\sqrt{C_k^{XX} \times C_k^{YY}}},
\end{equation}
for all pairs of $T$, $E$, and $B$.  The top row shows the spectra, 
with the fitting window denoted by vertical light-grey lines.   The
bottom row shows $\langle \rxy \rangle$, averaged over the fit window and
all frames,
as a function of sonic Mach number.  Error bars are found by first averaging
$r_k^{XY}$ over $k$ for each frame {within the fit window}, then taking the variance over frames.
Shown in horizontal grey lines are the observed values of \numvalue{$\rte=0.355$ and $\rtb=0.055$} \citep{Planck18XI}.

For the two even-parity modes, $T$ and $E$, we find significant positive correlation in all cases but one.  As the velocity decreases and magnetic 
field increases, $\rte$ increases well above the value observed by Planck. 
The general trend is of increasing correlation with increasing field 
strength.  For large sonic Mach numbers, correlations are more modest and mostly 
consistent with the observed Planck value of \numvalue{$\rte=0.35$}.  For 
large \mach, the effect of the magnetic field is not as pronounced as it is 
for lower \mach.  Earlier results, e.g.\ for $\alphae$ and the ratio of $E$ to $B$, show that the Planck data are consistent with $\mach>4$, and this is also compatible with what we see for $\langle \rte \rangle$.  We do observe a scale dependence in $\rte$ (rising toward small scales) that is not seen by Planck, which may be a sign that our simplified simulations are missing a key element of the ISM.

How do we draw conclusions from the $\langle \rte \rangle$ values, compared to
the Planck value of 0.35?  The Planck correlation looks most
compatible with slow velocities with moderate magnetic fields or fast velocities
with moderate-to-strong magnetic fields.  Earlier we saw that the $B/E$ power
ratio prefers fast velocities with $\mach > 4$, so the latter case seems to fit the bill.  Other combinations do not work: in our simulations, low $\mach$ and low $\alfmach$
(low velocity and strong magnetic field) would lead to a stronger correlation
than what Planck sees.  Low $\mach$ and high $\alfmach$ (low velocity and weak
magnetic field) leads to too small of a correlation, or a slight
anticorrelation.  We note that these simulations are idealized, and more realistic simulations may contain physical effects that reduce the filament alignment to the magnetic field and thus see the $TE$ correlations reduced.

It is hard to assess the $TB$ and $EB$ correlations, which we expect to be zero based on the physics in the simulation.  By sample variance, individual time snapshots can have a small nonzero correlation in certain wavenumber bands.  In the cross-correlation spectra (Fig.~\ref{fig.pearson}), which are averaged over 5\tdyn, the deviations from zero correlation are the same magnitude as the
mode-to-mode fluctuations in power, much smaller than the $TT$, $EE$, $BB$, and
$TE$ correlations, which we measure robustly.  Still, for $TB$ in the middle panel of the figure, some spectra appear to be mostly above zero for these realizations.  The error bars we draw on the mean $r_{TB}$ do intersect zero for most cases, but the fluctuations are smaller than we would expect compared to the size of our error bars, so maybe we have overestimated them.
We note that 14 of
the 21 cases have positive values, and this number or greater has 9\% cumulative 
probability in a binomial distribution with equal weight on positive and negative.  At fast velocities ($\mach > 4$), we find that 8 of 9 realizations are positive, which is 2\% probability.  Thus is not completely clear what to conclude.

We also show $r_{EB}$ in the third panel of the figure.
These correlations are on the same order as \rtb, and similarly have 14 of 21
simulations slightly positive, though consistent with zero according to our error prescription.  Above  $\mach > 4$, all 9 cases have positive correlation (0.2\% probability, but measured \textit{a posteriori}).

These tendencies toward positive parity-violating correlation values at high fluid velocity are somewhat puzzling
because all of the MHD physics we include respects parity.  The simulations
start with uniform density and are driven with a non-helical acceleration
pattern, and the base solver is an unsplit solver with no inherent asymmetry.
A larger, more systematic, and more resolved ensemble of simulations will be necessary to determine
 if we have inadvertently inserted some parity-violating effect, if this is simply a statistical fluctuation, or if there is a slight tendency for such MHD simulations to produce positive parity violations. 

\begin{table}
\begin{tabular}{lrrrr}
\hline
 Spectra & $a$ & $b$ & $c$ & $c/b$ \\
$\alpha_\rho$& -3.61& 0.16&  -0.00&  -0.03\\
$\alpha_v$& -3.86& 0.02&  0.14&  6.48\\
$\alpha_H$& -3.31& 0.02&  -0.28&  -18.13\\
$\alpha_{TT}$& -3.66& 0.15&  0.09&  0.62\\
$\alpha_{EE}$& -3.63& 0.17&  0.28&  1.60\\
$\alpha_{BB}$& -4.82& 0.28&  0.64&  2.28\\
\hline
\hline
\end{tabular}
\caption{Linear fits of the form $q = a + b \mach + c \alfmach$.}

\label{tab:multifit} \end{table}

\subsection{Linear fits and importance of parameters}
\label{sec.linear}

The nearly linear nature of the results in Figures \ref{fig.primspectra} and
\ref{fig.TEB_spectra} inspire us to fit the slopes of each of our quantities to a
linear relation of the form
\begin{align}
\alpha_q = a_q + b_q \mach + c_q \alfmach
\end{align}
where $q$ stands for density, velocity, magnetic field, $T$, $E$, and $B$.
These
fit coefficients are found in Table \ref{tab:multifit}.  
Noting that $b=\partial q/\partial \mach$ and
$c=\partial q/\partial \alfmach$,  we see that $b$ and $c$ give the relative importance of sonic
and \alf\ Mach numbers on each quantity.  The third column of Table
\ref{tab:multifit} gives the ratio of $c/b$, which denotes the relative impact
of the two.  Density, $\rho$, depends only on
sonic Mach number $(b_\rho=0)$ while the velocity spectrum is more influenced by
\alfmach\ ($c_\rho/b_\rho=6.5$).
Sonic Mach number determines $\alpha_{TT}$, while \alf\ Mach number determines
$\alpha_{BB}$.  

From this linear process, we can derive an typical \mach\ and \alfmach\ for
the ISM.  By simultaneously
solving the linear equations for \numvalue{$\alpha_{\rm{EE}}=-2.4$} and
\numvalue{$\alpha_{\rm{BB}}=-2.5$,} we
find an ideal \numvalue{$\mach=4.7$ and $\alfmach=1.5$} from the slopes.  {This combination would produce an appropriate $B/E$ power ratio, but would probably underproduce \rte, which would prefer a somewhat smaller $\alfmach \sim 0.7$ or so and a higher velocity $\mach \sim 5$--$6$ to compensate the slope.}  Of course, the true values for
\mach\ and \alfmach\ may vary substantially from point to point in the sky, as the
ISM is a multiphase medium and the sound speed and kinetic energy are determined
by the phase.  However these give a typical value for reproducing the
geometrical structures in the ISM.

\subsection{Parallel versus perpendicular projections}
\label{sec.xhat}

We focus primarily on the behavior of projections perpendicular to the mean magnetic
field because it is a more physically appropriate configuration to compare to the sky.
To observe a signal comparable to projecting our boxes along the mean field
would require the field to be radially directed away from the earth, and to be
coherent over a large fraction of the optical depth of the ISM.  This is
unlikely.  However, the real signal will be an admixture of orientations along
the line of sight, so we present the major differences here.

\begin{figure} \begin{center}
\includegraphics[width=\linewidth]{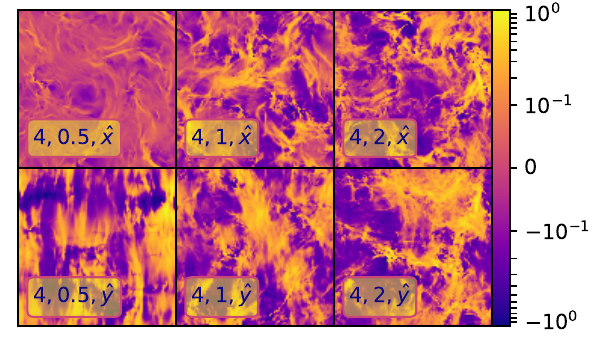}
\caption{Projections of $E$ for the sonic mach $\mach=4$ simulations, with targeted $\alfmach = 0.5, 1, 2$.
\emph{(Top row)} projection along $\hat{x}$, with the magnetic field coming out of
the page and \emph{(Bottom row)} projection along $\hat{y}$, with the field
pointing up (as in Fig.~\ref{fig.proj}).  Field strength increases to the left.}
\label{fig.proj_xhat} \end{center} \end{figure}

Figure \ref{fig.proj_xhat} shows projections of the $\mach=4$ suite in the
$\hat{x}$ direction, along the magnetic field (top row), and the $\hat{y}$
direction, perpendicular to the field (bottom row).  Magnetic field strength increases to the
left.  The impact of the field is most apparent for the $\alfmach=0.5$
simulations with the strongest magnetic field.  The mean magnetic field is out of the page in the top row, and
suppresses motion across the line of sight, while the mean field is vertical in
the bottom row, and suppresses motion in the horizontal direction.

\begin{figure*} \begin{center}
\includegraphics[width=\linewidth]{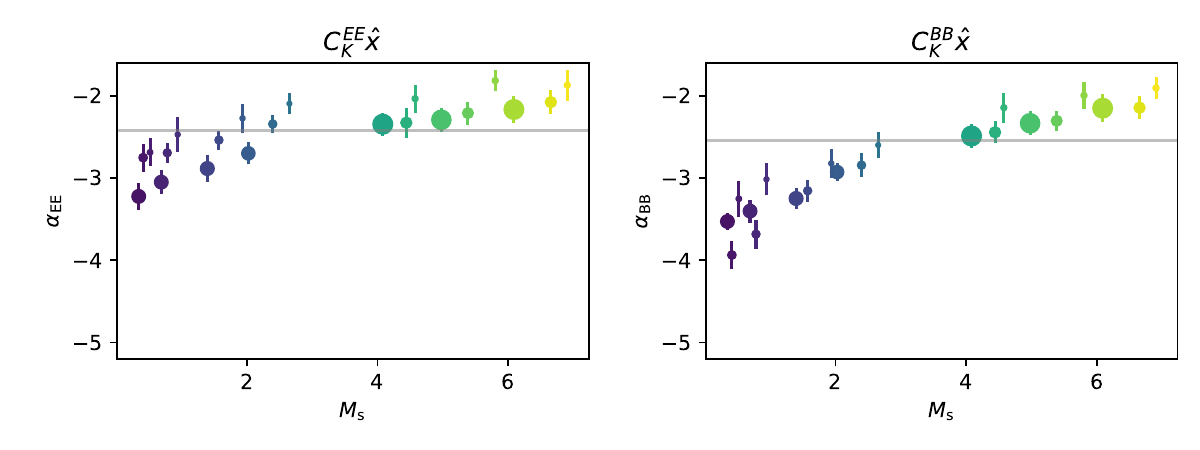}
\includegraphics[width=\linewidth]{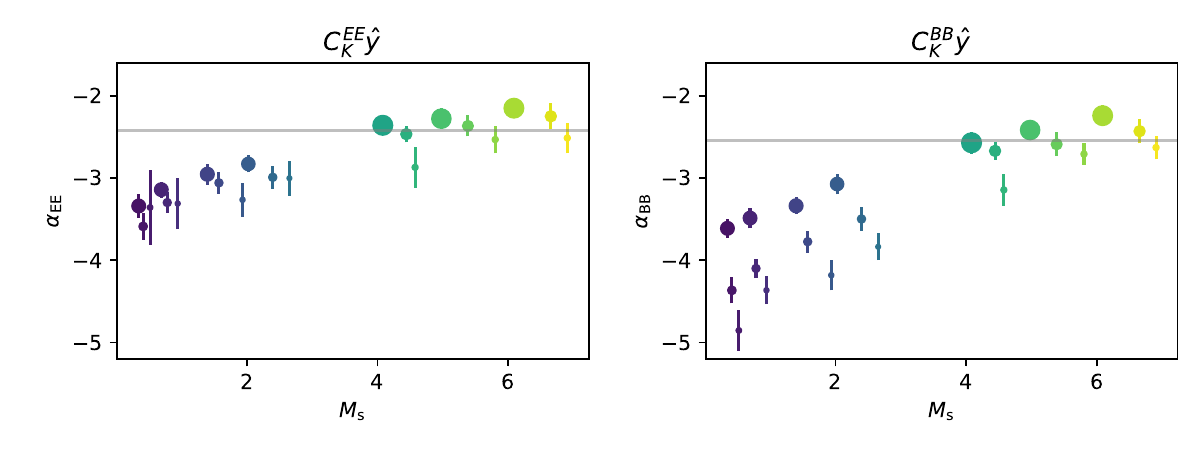}
\caption{Amplitudes for $C_k^{EE}$ and $C_k^{BB}$ along the magnetic field ($\hat{x}$, top row) and
across the field ($\hat{y}$, bottom row), as in Fig.~\ref{fig.TEB_slopes}.   }
\label{fig.slopes_xhat} \end{center} \end{figure*}

Figure \ref{fig.slopes_xhat} shows $\alphae$ and $\alphab$ vs $\mach$ for
parallel projections (top row) and perpendicular projections (bottom row, same
as Figure \ref{fig.TEB_slopes}, reproduced for ease of comparison).  For
the weakly magnetized cases (large points) the behavior is comparable between
the two directions, as
expected.  For the more strongly magnetized case, increasing magnetic field has
the opposite effect on the slope between the two directions.  For the $\hat{x}$
projection, increasing magnetic field makes $\alphae$ and $\alphab$
slightly more shallow.  For the parallel direction, increase mean field causes $\alphae$
to become steeper, but $\alphab$ steepens more dramatically.  

\begin{figure*} \begin{center}
\includegraphics[width=\linewidth]{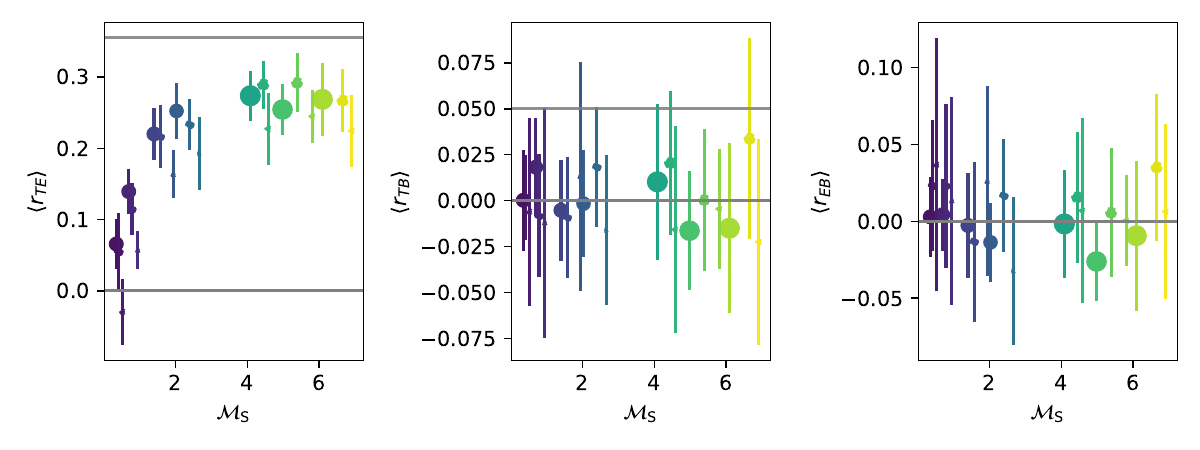}
\caption{Average correlations $r_{TE}$, $r_{TB}$ and $r_{EB}$, left to right
respectively, for the $\hat{x}$ projections along the magnetic field. }
\label{fig.r_xhat} \end{center} \end{figure*}

Figure \ref{fig.r_xhat} shows the cross correlation, $r_{TE}$, $r_{TB}$ and
$r_{EB}$ for the perpendicular projections.  The $TE$ correlation increases
with \mach\ to a typical value of about 0.25, slightly smaller than the value of
0.355 observed on the sky.  The correlations with $B$ look consistent with zero for the parallel projections.


\section{Discussion}
\label{sec.discussion}

Here we discuss our results in the context of our model foregrounds as an
ensemble of filaments.
In \citet{Huffenberger20}, we model $E$ and $B$ with filaments that have an
aspect ratio $\epsilon\in[0,1]$ threaded by magnetic fields at an angle,
$\theta_{LH}$.  
As $\epsilon$ increases, making the filaments more round, the ratio of $B$ to
$E$ increases as a shorter, rounder filaments have proportionally less $E$ (see
Figure 6 of \citealt{Huffenberger20}).  This also
explains
the decrease in $r^{TE}$ as $\epsilon$ increases.  These predictions from the
filament model are
consistent with our findings with the turbulent boxes.  As \mach\ and
\alfmach\ increase, the ability of the magnetic field to suppress instability decreases,
and shorter filaments are expected.  This can be seen in projections and in the
power spectra of $\rho.$  Commensurate, we find an increase in $B/E$ and a
decrease in $r^{TE}$ with increasing \mach.  Again, the shorter structures as a
result of increased \mach\ and \alfmach\ have effects on the polarization power spectra that are in agreement, at least
qualitatively, with the model of \citet{Huffenberger20}.  
\citet{Clark21}  model the parity violating correlation as a misalignment  between filamentary structure and magnetic field direction in a similar filamentary
framework, and compare to simulations.
In future, we will examine
filamentary properties of these cubes to further explore the predictive power of
\citet{Huffenberger20} and \citet{Clark21}.

\section{Conclusions}
\label{sec.conclusions}

In this work, we examine the $E-$mode and $B-$mode spectra from a suite of 
idealized, magnetized, and turbulent simulations.  We find that isothermal turbulence alone is
enough to reproduce the observed values of $\alpha_{EE}$ and $\alpha_{BB}$, as
well as the ratio of amplitudes, $A_{BB}/A_{EE}$, for suitable values of sonic Mach
number, \mach, and \alf\ Mach number, \alfmach.   We additionally find that
the observed correlation of $T$ and $E$, \numvalue{$r^{TE}=0.3$}, is naturally
reproduced by the turbulence at high \mach\ and an appropriate magnetic field strength.  Parity-violating correlations with $B$ are spectrally flat and near
zero, certainly below $0.05$, but the results are somewhat murky. 
We suggest that a ``typical'' patch of the sky has $\mach=4.7$, $\alfmach=1.5$,
based on linear interpolation of the $E$- and $B$-mode slopes, but $\rte$ prefers lower \alfmach\ and higher \mach.  

The density spectrum is found to be tightly related to \mach, with slope
\numvalue{$\alpha_\rho\sim -3.6 +0.16 \mach $.  }  This is due to the fact that shock
thickness decreases with \mach, leading to smaller scale structure and faster
growth of instabilities such as Richmeyer-Meshkov and Rayleigh Taylor.
The velocity spectra is relatively insensitive to \mach\ for $\mach>2$.
Supersonic slope values cluster around $\alpha_v\sim-3.5$ for super-\alf\
values, slightly shallower than the Kolmogorov value of $-11/3$.  
Magnetic spectral slopes are relatively insensitive to \mach, and decrease with
decreasing magnetic field.

The projected quantities, $T$, $E$ and $B$, also depend on \mach\ and \alfmach.
In these perfectly optically thin models, $T$ is the integral of $\rho$ along
the line of sight, and it is found that $\alpha_T\sim\alpha_\rho$ with some small
dependence due to the magnetic field.  
$E$ is found to depend on both \mach\ and \alfmach, from
\numvalue{$\alpha^{EE}\in[-3.5,-2]$ and $\alpha^{BB}\in[-4.5,-2.2]$.  }

{In future simulation studies, we need higher resolution to increase our inertial range and improve our accuracy, particularly on the slopes.  We also need larger statistical ensembles to quantify the sample variance in the $TB$ and $EB$ correlations.  These parity-violating correlations are important for measurements of detector calibration, gravitational lensing, and cosmic birefringence.}

\section*{Acknowledgements}
\addcontentsline{toc}{section}{Acknowledgements}

Support for this work was provided in part by the National Science Foundation
under grants AST-1616026 and AST-2009870, NASA under grant NNX17AF87G and 80NSSC23K0466, and the DOE under grant DE-SC0024462.  Simulations were performed on \emph{Stampede2}, part of the Extreme Science and Engineering Discovery Environment \citep[XSEDE;][]{Towns14}, which is supported by National Science Foundation grant number
ACI-1548562, under XSEDE allocation TG-AST140008.  We thank the anonymous referee for several suggestions to clarify the text.

\bibliographystyle{mnras}
\bibliography{apj-jour,b2}  

\end{document}